\documentclass{PoS}

\usepackage{amsmath}
\usepackage{wrapfig}
\usepackage{caption}
\usepackage{subcaption}

\title{Simulations of gaussian systems in Minkowski time}

\ShortTitle{Simulations of gaussian systems in Minkowski time}

\author{\speaker{Blazej Ruba}\thanks{This work is supported in part by the NCN grant: UMO-2016/21/B/ST2/01492.}\\
        Jagiellonian University\\
        E-mail: \email{blazej.ruba@doctoral.uj.edu.pl}}

\author{Jacek Wosiek\\
        Jagiellonian University\\
        E-mail: \email{jacek.wosiek@uj.edu.pl}}

\abstract{Many research programs aiming to deal with the sign problem were proposed since the advent of lattice field theory. Several of these try to achieve this by exploiting properties of analytic functions. This is also the case for our study. There auxiliary complex variables are introduced and desired weight is obtained after integrating them out. In this note we clarify conceptual difficulties with this procedure encountered in previous works. In the process we observe an exciting connection with thimbles and discover an interesting hidden symmetry present in the problem. Problem of negative eigenvalues of the action will be revisited and considered from a different perspective. As a byproduct we perform simulations of simple quantum systems directly in Minkowski time.}

\FullConference{The 36th Annual International Symposium on Lattice Field Theory - LATTICE2018\\
		22-28 July, 2018\\
		Michigan State University, East Lansing, Michigan, USA.}

\begin{document}

\section{Introduction}
Our purpose is to numerically compute averages with respect to a holomorphic weight $\rho$:
\begin{equation}
    \langle \mathcal O \rangle_{\rho} = \frac{\int_{\Gamma} d z \ \rho(z) \mathcal O (z) }{\int_{\Gamma} d z \ \rho(z)},
    \label{eq:Rho_Average}
\end{equation}
where $z=(z^1,...,z^n)$ is a complex vector, $d z = d z^1 ... d z^n$ and $\Gamma \subseteq \mathbb C^n$ is an $n$-dimensional surface without a boundary (from now on called an $n$-cycle), typically $\Gamma = \mathbb R^n$. If $\left. \rho \right|_{\mathbb R^n}$ is complex and $n$ is large, this problem is intractable. A possible way out is to find a positive representation for $\rho$, i.e. a positive function $P$ on $\mathbb R^{2n}$ such that for analytic $\mathcal O$ sufficiently well behaved at infinity
\begin{equation}
\langle \mathcal O \rangle_{\rho} =   \langle \mathcal O \rangle_P := \frac{\int_{\mathbb R^{2n}} d x d y \ P(x,y) \mathcal O (x+iy)}{\int_{\mathbb R^{2n}} d x d y \ P(x,y)} .
  \label{eq:P_avg}
\end{equation}
It is known \cite{Salcedo, Weingarten, RubaWyrzyk} that infinite families of positive representations exist under mild assumptions on $\rho$. Some suggestions how to find them were put forward \cite{Parisi, Klauder, Pham, Cristoforetti}. In this work we develop the Beyond Complex Langevin (BCL) approach proposed in \cite{Wosiek_zzbar}.

We assume existence of a positive distribution $P$ which is real analytic. Then it may be extended uniquely to complex $x,y$. We introduce coordinates $z=x+iy$ and $\overline z= x - i y$. Notice that $\overline z$ is equal to $z^*$ - the complex conjugate of $z$ - only on the subspace $\mathbb R^{2n} \subseteq \mathbb C^{2n}$. Differential $2n$-form $P(z, \overline{z}) dz d \overline{z}$ is closed\footnote{This is equivalent to Cauchy-Riemann equations.}, so Stokes theorem may be used to replace $\mathbb R^{2n}$ by a more convenient integration surface. Suppose that there exist $n$-cycles $\Gamma \subseteq \{ z \}, \overline{\Gamma} \subseteq \{ \overline z \}$ such that P is rapidly decreasing on $\Sigma = \Gamma \times \overline{\Gamma}$. Let $\mathcal B(\Lambda) = \{ (z, \overline z) \in \mathbb C^{2n} | |z|^2 + |\overline z|^2 \leq \Lambda^2 \}$ and choose a tube $\mathcal T(\Lambda)$ which connects $\mathcal B(\Lambda) \cap \mathbb R^{2n}$ with $\mathcal B(\Lambda) \cap \Sigma$, so that $\mathcal C(\Lambda) = \mathcal T(\Lambda) \cup \mathcal B(\Lambda) \cap \left( \mathbb R^n \cup \Sigma \right)$ is a closed cylinder. Then
\begin{equation}
\int_{\mathcal C(\Lambda)} d z d \overline z \ P(z, \overline z) \mathcal O(z)   =0.
\end{equation}
It follows that the integrals over $\mathbb R^n$ and $\Sigma$ are equal, provided that the integral over $\mathcal T(\Lambda)$ tends to zero as $\Lambda \to \infty$. In this case a sufficient condition for (\ref{eq:P_avg}) is given by
\begin{equation}
\int_{\bar \Gamma} d \overline z \ P(z, \overline z) = \mathcal N \rho(z).
\label{eq:zbar_integration}
\end{equation}

\section{Single variable}

In \cite{Wosiek_zzbar} a positive representation $\rho=e^{-\sigma z^2}$ was found as a solution to (\ref{eq:zbar_integration}). One may take $P(z, \overline z)=e^{-S_{\mathrm{BCL}}(z, \overline z)}$ with an action $S_{\mathrm{BCL}}=a^* z^2 + 2b z \overline z + a \overline z ^2$, where $a,b$ are given in terms of $\sigma$ and one free parameter - a hyperbolic angle $\chi \in (0, \infty)$:
\begin{subequations}
\begin{gather}
a =  - \sigma ^* \sinh^2 \left( \chi \right), \\
b =  \frac{| \sigma |}{2} \sinh(2\chi),
\label{eq:1dgauss_bare_parametrization}
\end{gather}
\end{subequations}
Choice $\sinh \left( \chi \right) = \left| \frac{\mathrm{Re} \ \sigma}{\mathrm{Im} \ \sigma} \right|$ reproduces the result obtained in (\cite{AmbjornYang}) using the complex Langevin approach. Limit $\chi \to \infty$ is even more interesting. For $\overline{z}=z^*$ action takes the form
\begin{equation}
S_{\mathrm{BCL}}(z, z^*) = 4 \sinh^2 \left( \chi \right) \left( \mathrm{Im} \ \sqrt{\sigma} z \right)^2 + |\sigma z^2 | + O \left( e^{- 2\chi} \right).
\end{equation}
It follows that asymptotically $P$ is concentrated on the line $\mathfrak L =\{ z \in \mathbb C \ | \  \mathrm{Im}(\sqrt{\sigma z^2}) = 0 \}$,
\begin{equation}
   \left. P(z, \bar z) \right|_{\bar z = z^*} \xrightarrow{\chi \to +\infty} \mathcal N \delta \left( \mathrm{Im} \sqrt{\sigma z^2}  \right) \rho(z),
  \label{eq:gaussian_thimble_concentration}
\end{equation}
This is the original weight $\rho$ with the integration contour rotated so that the phase of $\rho$ is constant.


Equation (\ref{eq:gaussian_thimble_concentration}) could be attributed to the fact that the action we have considered is quadratic. We will see later that it remains true also for the action $S_{\mathrm{BCL}}= (a^* z^2 + 2b z \bar z + a \bar z ^2)^2$. In this case
\begin{equation}
\rho(z) = \int_{\sqrt{a^*} \mathbb R} d \overline z \ e^{-S_{\mathrm{BCL}}(z, \overline z)} =  \mathcal N \sqrt{ \sigma z^2} e^{- \frac{1}{2} \sigma^2 z^4} K_{\frac{1}{4}} \left( \frac{\sigma^2 z^4}{2} \right),
\label{eq:quartic_rho}
\end{equation}
One of the new results is the calculation \cite{Thesis} of moments
\begin{equation}
    \langle z^{2k} \rangle_{\rho(z)} = \langle z^{2k} \rangle_{P(z, \bar z)} = \frac{1}{\sqrt{\pi}} \frac{\Gamma \left( \frac{2k+1}{2} \right)}{ \Gamma \left( \frac{k+2}{2} \right)} \left( \frac{1}{2\sigma} \right)^k.
\end{equation}
$\langle z^{2k} \rangle_{\rho}$ and $\langle z^{2k} \rangle_{P}$ were computed independently, providing a nontrivial consistency check.

Just as in the previous example, $\rho(z)$ doesn't depend on $\chi$. This may be traced back to the fact that in both cases $S_{\mathrm{BCL}}$ is annhilated by the commuting differential operators
\begin{subequations}
\begin{gather}
\overline D = \sinh(\chi) \frac{\partial}{\partial \chi} + \left( \frac{|\sigma| z}{\sigma} \sinh{\chi} - \cosh \left( \chi \right) \overline z \right) \frac{\partial}{\partial \overline z}, \\
D = \sinh(\chi) \frac{\partial}{\partial \chi} + \left( \frac{|\sigma| \overline z}{\sigma^*} \sinh(\chi) - \cosh(\chi) z \right) \frac{\partial}{ \partial z}.
\end{gather}
\end{subequations}

The limit $\chi \to \infty$ may be analyzed using the saddle point method. It turns out that the gradient of $S_{\mathrm{BCL}}$ vanishes on $\mathfrak L$ asymptotically for $\chi \to \infty$. This set may be parametrized as $\mathfrak L = \{ x+ iy_*(x) \ | \ x \in \mathbb R \}$, where $y_{*}(x) = - \frac{\mathrm{Im} \ \sigma}{|\sigma|+ \mathrm{Re} \ \sigma} x$. For any test function $f$ the integral $\int_{\mathbb R^2} dx dy \ P(x,y)  f(x,y)$ may be simplified by replacing $f(x,y)$ with $f(x, y_{*}(x))$. The only dependence on $y$ is then that of the weight $P(x,y)$ itself. In contrast to the standard saddle point method, integration $\int dy \ P(x,y)$ may not be performed by expanding $S_{\mathrm{BCL}}$ to the second order in $y - y_*(x)$. The problem is that the second derivatives of $S$ vanish at $0$. It is necessary to keep the quartic terms. The final result is
\begin{equation}
\int_{\mathbb R} d y \ P(x,y) = \mathcal N \rho(x+i y_*(x)).
\end{equation}
This completes the proof of (\ref{eq:gaussian_thimble_concentration}) for the quartic action. Results of this section suggests that there might exist a connection between our approach and the Lefschetz thimble method.


\section{Quantum mechanics}

We will now discuss positive representations for simple Feynman weights in Minkowski time. Following \cite{Wosiek_zzbar} we take a simple ansatz for the BCL action:
\begin{equation}
S_{\mathrm{BCL}}(z, \overline z) = \sum_{j=1}^n \left[ i \beta \overline z_j^2 - i \beta z_j^2 +2b \overline z_j z_j + 2 \gamma \overline z_j z_{j+1} + 2 \gamma \overline z_{j+1} z_j \right],
\label{eq:zzbar_action}
\end{equation}
where $z_{n+1}=z_1$, $\overline z_{n+1}=z_1$. Eigenvalues of $S_{\mathrm{BCL}}$ may be found by Fourier transformation, which reduces the problem to diagonalization of $2 \times 2$ and $4 \times 4$ matrices. The final result is
\begin{equation}
\lambda_{k \pm} = 2b + 4 \gamma \cos \left( \frac{2 \pi k}{n} \right) \pm 2 |\beta|, \quad  k=0,...,n-1.
\end{equation}
Eigenvalues $\lambda_{0 \pm}$ are nondegenerate. For each $k \neq 0$ there are two linearly independent eigenvectors to eigenvalues $\lambda_{k +}$ and $\lambda_{k -}$. Weight $\rho=e^{iS_{\mathrm{phys}}}$ is obtained by integrating out $\overline z$ variables:
\begin{equation}
S_{\mathrm{phys}}(z) = -i \log \int d \overline z \ e^{- S_{\mathrm{BCL}}} =\sum_{j=1}^n \left[ \frac{2b \gamma}{\beta} (z_{j+1}-z_j)^2 + \frac{4 \gamma^2}{\beta} \left( \frac{z_{j+2}-z_j}{2} \right)^2 - \frac{(b+2 \gamma)^2 - \beta^2}{\beta} z_j^2 \right].
\end{equation}
Demanding that it coincides (up to an unusual discretization of the kinetic term) with the action of harmonic oscillator with angular frequency $\omega$ and time step $\epsilon$ we get the following conditions:
\begin{subequations}
\begin{gather}
\frac{2 \gamma (b+ 2 \gamma)}{\beta} = \frac{m}{2 \epsilon}, \\
\frac{(b+2 \gamma)^2 - \beta^2}{\beta} = \frac{m \omega^2 \epsilon}{2}.
\end{gather}
\end{subequations}
One can show that these equations are in contradiction with all $\lambda_{k \pm}$ being positive. We will address this issue separately for the free particle ($\omega=0$) and for the harmonic oscillator.

\subsection{Free particle}

For $\omega=0$ and negative $\beta$, BCL action (\ref{eq:zzbar_action}) may be rewritten as
\begin{equation}
S_{\mathrm{BCL}}= \sum_{j=1}^n \left[ \frac{m}{2 \epsilon} ( \overline z_{j+1} - \overline z_j ) (z_{j+1}-z_j)  + 2 |\beta| ( x_j - y_j )^2  \right].
\label{eq:free_particle_action}
\end{equation}
In this case $\lambda_{0-}=0$, but other eigenvalues are positive. Existence of the zero mode reflects the translational symmetry of the system. It may be removed by changing the boundary conditions from periodic to Dirichlet. It was shown in \cite{Wosiek_zzbar} that $e^{-S_{\mathrm{BCL}}}$ becomes a positive representation for the Feynman weight if the limit $\beta \to - \infty$ is taken before the continuum limit. This removes unusual terms from the acton $S_{\mathrm{phys}}$. Formula (\ref{eq:free_particle_action}) show that in this limit $P(z, z^*)$ becomes concentrated on the subspace $ \{ x + i y \in \mathbb C^n \ | \ \forall j : \ x_j = y_j \in \mathbb R \}$, so we have
\begin{equation}
P \xrightarrow{\beta \to - \infty} \mathcal N \prod_{j=1}^n \delta(x_j - y_j) e^{ \frac{i m}{2 \epsilon} (z_{j+1}-z_j)^2}.
\label{eq:free_particle_large_beta}
\end{equation}
This is the standard Feynman weight with contours of all $z_j$ rotated by $\frac{\pi}{4}$.

It turns out that taking the limit $\beta \to - \infty$ is not necessary. Properties of $S$ at finite $\beta$ are easiest to understand if the following change of variables is made:
\begin{subequations}
\begin{gather}
    u_j = \frac{1}{\sqrt{2}} (x_j + y_j), \\
    v_j =  \frac{1}{\sqrt{2}} (x_j - y_j).
    \end{gather}
\end{subequations}
Then the action takes the form
\begin{equation}
S_{\mathrm{BCL}} = \sum_{j=1}^n \left( \frac{m}{2 \epsilon} \left( u_{j+1} - u_j \right)^2 + \frac{m}{2 \epsilon} (v_{j+1}-v_j)^2 + 4 |\beta| v_j^2 \right).
\end{equation}
It describes a free particle $u_j$ and a spurious degree of freedom $v_j$. Correlation function for $v_j$ at large $n$ may be found via contour integration or strong coupling expansion. The result is
\begin{equation}
    \langle v_{j+d} v_j \rangle \approx \frac{ e^{- \frac{d \epsilon}{\xi}}}{8\sqrt{ |\beta| \left(  |\beta| + \frac{m}{2\epsilon} \right)}},
\end{equation}
with errors vanishing for $n \to \infty$. Correlation length is equal to $\xi = \frac{\epsilon}{2 \  \mathrm{arsinh} \sqrt{\frac{2 |\beta| \epsilon}{m}}}$. Therefore $v_j$ decouples in the continuum limit if and only if $\beta$ changes with $\epsilon$ in such a way that
\begin{equation}
\lim_{\epsilon \to 0} \frac{\beta}{\epsilon}= - \infty.
\end{equation}
This condition is much weaker than requirement that $\beta \to - \infty$ at fixed $\epsilon$.

\subsection{Harmonic oscillator}

It was observed already in \cite{Wosiek_zzbar} that what we did for the free particle can't be repeated verbatim for the harmonic oscillator. The problem is that some of the eigenvalues $\lambda_{k \pm}$ become negative. We will now show that this phenomenon is a feature of the system, rather than peculiarity of our approach. Consider the Minkowski time action
\begin{equation}
S = \sum_{j=1}^n \left[ \frac{m}{2 \epsilon} (x_{j+1}-x_j)^2 - \frac{m \omega^2 \epsilon}{2} x_j^2 \right].
\end{equation}
We change variables to Fourier amplitudes
\begin{equation}
x_j = \frac{1}{\sqrt{n}} \sum_{k } \tilde x_k e^{\frac{2 \pi i kj}{n}}.
\end{equation}
The action then takes the form
\begin{equation}
S = \sum_k \left[ \frac{2m}{\epsilon} \sin^2 \frac{k \pi}{n} - \frac{m \omega^2 \epsilon}{2} \right] \tilde x_k \tilde x_{-k}.
\end{equation}
For small $T=n \epsilon$ there is only one negative eigenvalue, corresponding to variable $\tilde x_0$. One can get rid of it by imposing Dirichlet boundary condition. Two additional eigenvalues become negative after each oscillation period $\frac{2 \pi}{\omega}$. These can't be eliminated so easily. Presence of eigenvalues with different signs means, that to obtain postive and normalizable weight one has to rotate contours of integration by a $k$-dependent angle: 
\begin{equation}
    \tilde x_k = \begin{cases}
    e^{-\frac{i \pi}{4}} \ \tilde q_k & \mbox{for } |k| \leq \kappa, \\
    e^{\frac{i \pi}{4}} \ \tilde q_k & \mbox{for } |k| > \kappa,
    \end{cases}
    \label{eq:QHO_rotations}
\end{equation}
where $\kappa = \left \lfloor{ \frac{\omega T}{2 \pi}}\right \rfloor$. Then we perform inverse Fourier transformation:
\begin{equation}
q_j = \frac{1}{\sqrt{n}} \sum_{k=-K}^K \tilde q_k e^{\frac{2 \pi i k j}{n}}.
\label{eq:QHO_qvar_Fourier}
\end{equation}
In new variables action takes the form $S=S_1 + S_2$, where
\begin{subequations}
\begin{gather}
    S_1(q) = i \sum_{j=1}^n \left[ \frac{m}{2 \epsilon} (q_{j+1}-q_j)^2 - \frac{m \omega^2 \epsilon}{2} q_j^2 \right], \\
    S_2(q) = i \frac{m}{T} \sum_{k=-\kappa}^{\kappa} \left( \omega^2 \epsilon^2 - 4 \sin^2 \frac{k \pi}{n} \right) \left| \sum_{j=1}^n q_j e^{\frac{2 \pi i k j}{n}} \right|^2.
\end{gather}
\label{eq:real_time_QHO}
\end{subequations}
\begin{wrapfigure}{R}{0.5\textwidth}
    \centering
    \includegraphics[width=0.45\textwidth]{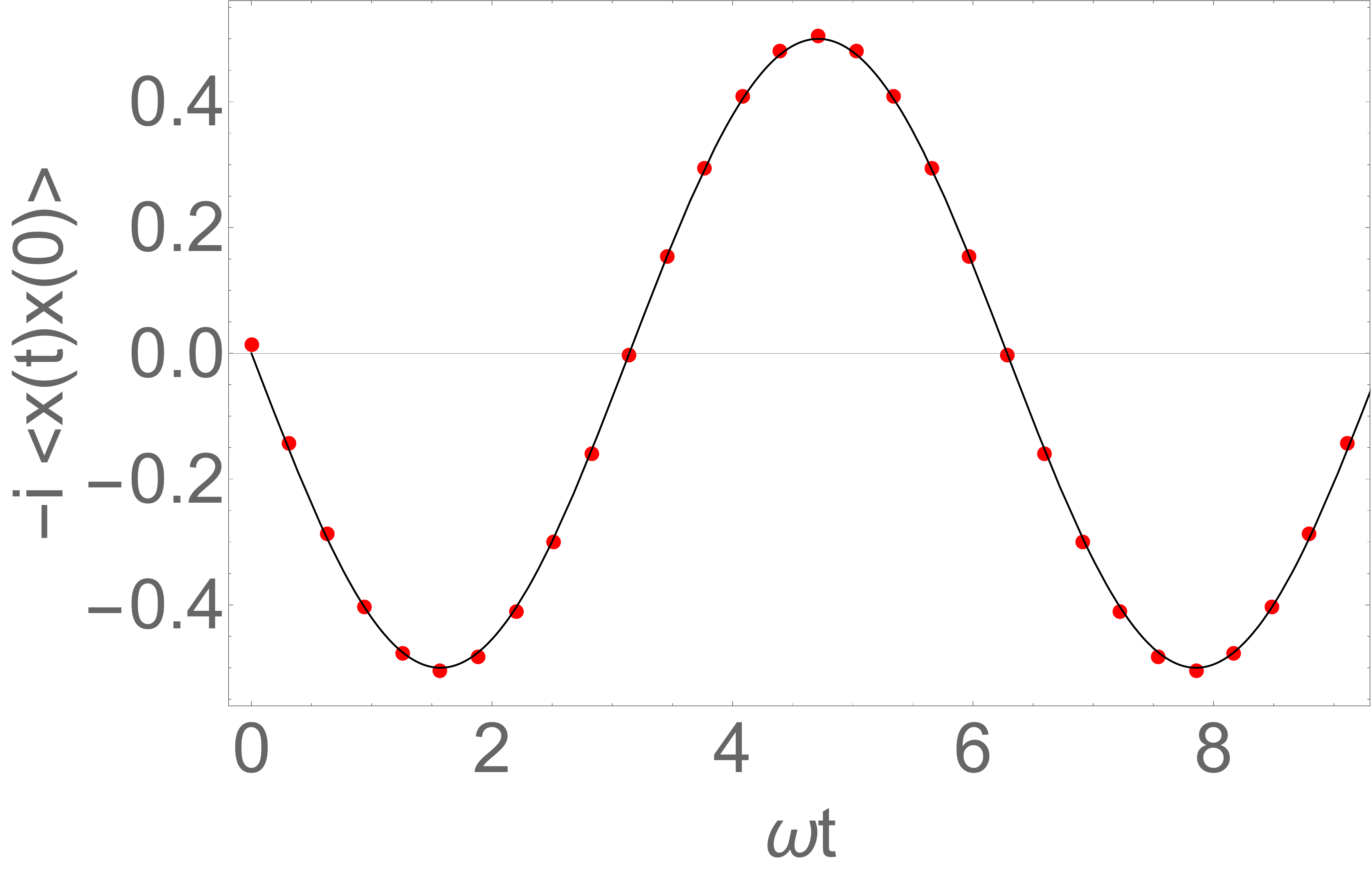}
    \caption{\label{fig:QHO_Mink_Correlator}Two point function $\frac{\mathrm{tr} \left( e^{-iT \hat H} \hat x(t) \hat x \right)}{\mathrm{tr} \left( e^{-iT \hat H} \right)}$, $t \in [0,T]$ for the harmonic oscillator calculated in a Monte Carlo simulation. Black line is the exact result. Uncertainties are too small to be seen on this plot.}
\end{wrapfigure}
The first term is up to $i$ factor the standard action for harmonic oscillator in Minkowski time with $x_j$ replaced by $q_j$. Due to the presence of $-$ potential term in $e^{iS_1}$ blows up at infinity. Second term stabilizes this divergence. For real $q_j$ distribution $e^{iS(q)}$ is positive and normalizable. By construction it is equivalent to the Feynman weight. In fact this is the exact thimble decomposition. Since $x=0$ is the only saddle point, there is only one thimble. The price to pay is introduction of nonlocal terms in the action and in the relation between $x_j$ and $q_j$. Two point function calculated by Monte Carlo simulation of the action (\ref{eq:real_time_QHO}) is presented on Figure \ref{fig:QHO_Mink_Correlator}.


There is an interesting interpretation of the negative eigenvalues of $S(x)$. Classical Morse theorem \cite{Duistermaat} states that the second variation of the action with the endpoints fixed is positive-definite for sufficiently small times and develops a negative eigenvalue each time a focal point is crossed. Purely quadratic actions are equal to their second variation, so information about the eigenvalues of $S(x)$ is obtained from this proposition. In the case of periodic boundary conditions one has to take into account the fact that there is one more variable integrated over - namely the coordinate of the endpoint. Thus the number of negative eigenvalues of $S$ with periodic boundary conditions is either equal or greater by one than then number of negative eigenvalues with Dirichlet boundary conditions. This is in agreement with our explicit calculations. The upshot is that existence of caustics seems to be an obstruction to the existence of local positive representations.

\section{Quantum fields}

\begin{figure}[th]
\centering
\begin{subfigure}{.5\textwidth}
  \centering
  \includegraphics[width=.5\linewidth]{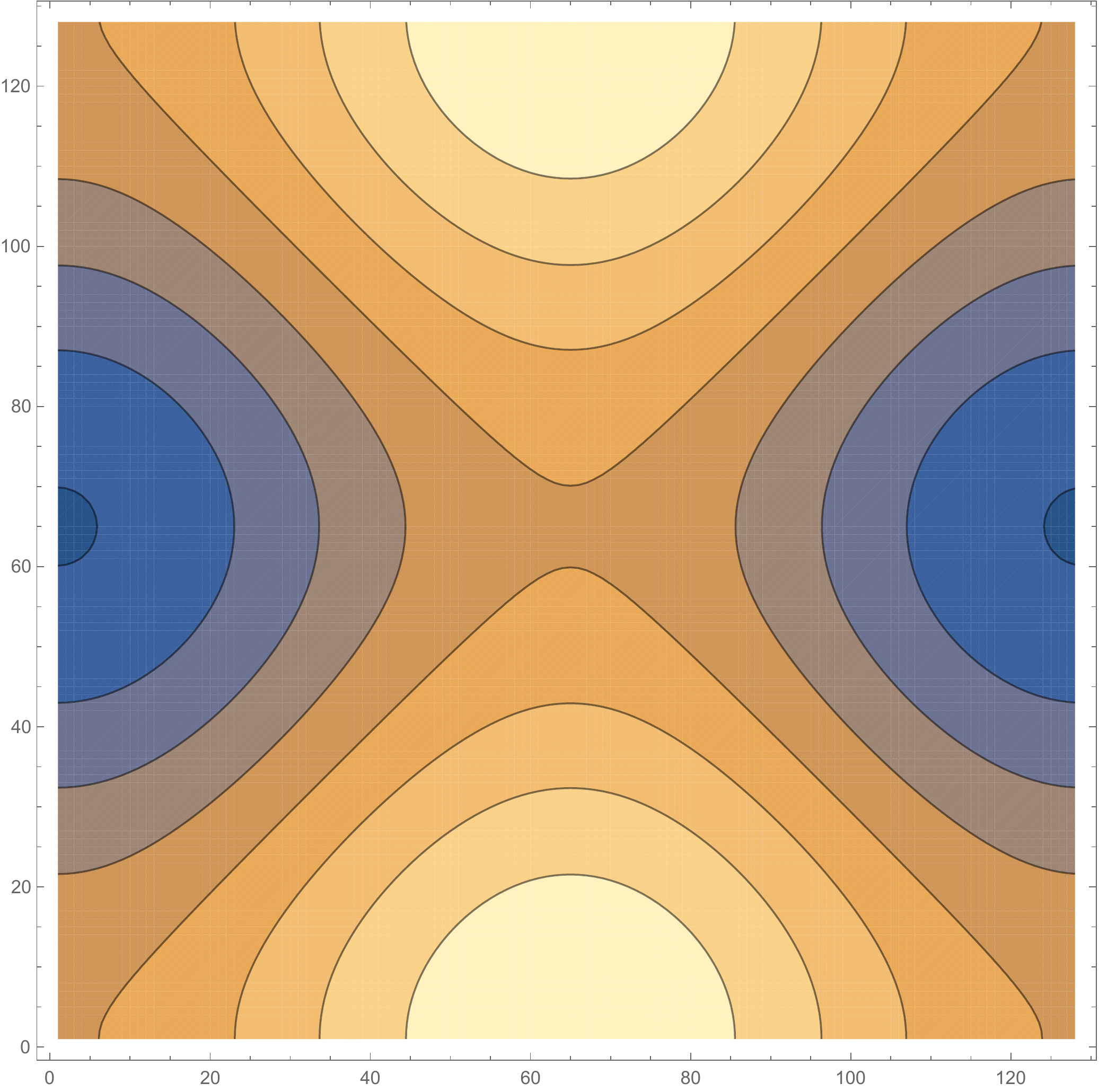}
  \label{fig:sub1}
\end{subfigure}%
\begin{subfigure}{.5\textwidth}
  \centering
  \includegraphics[width=.5\linewidth]{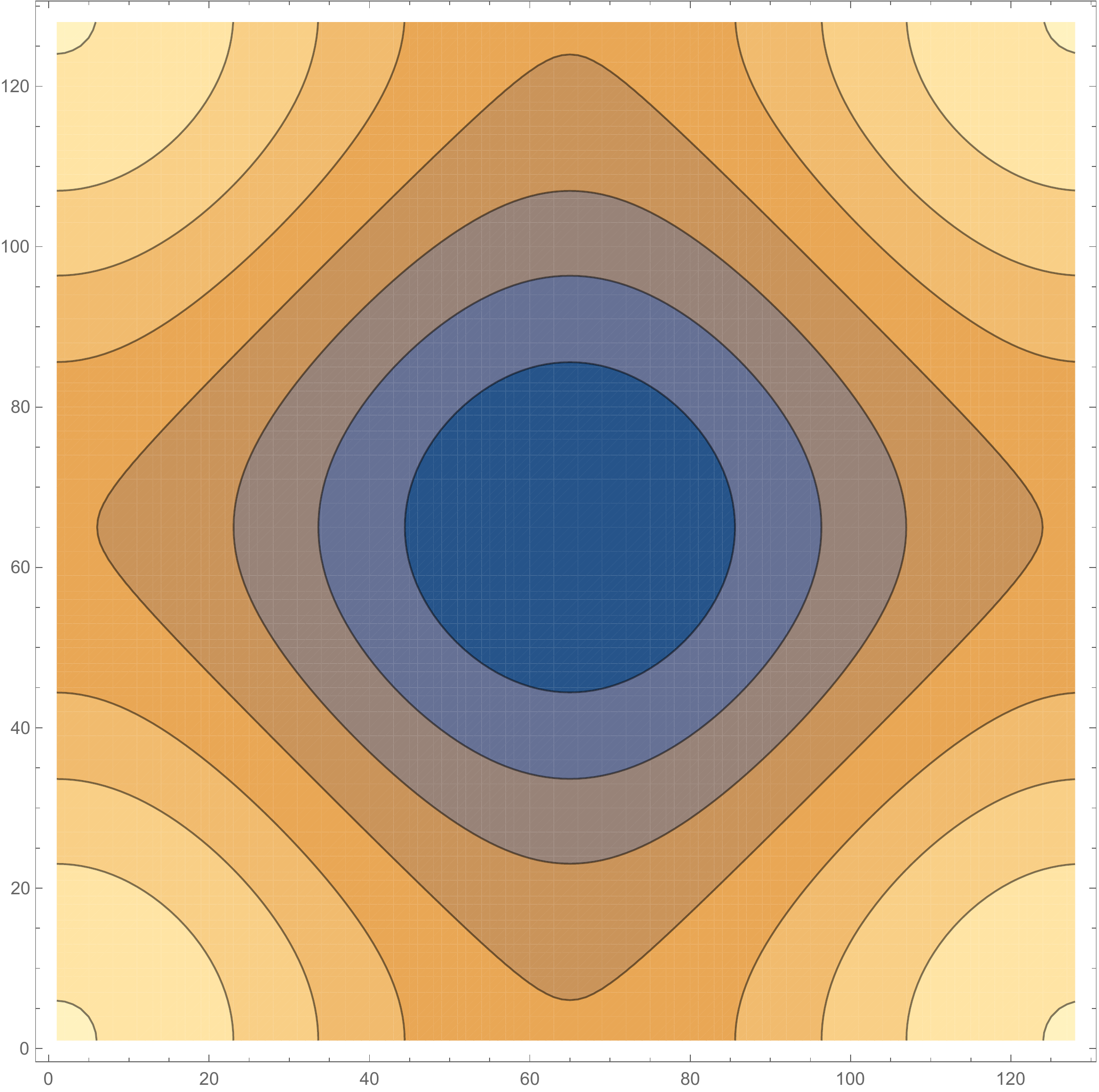}
  \label{fig:sub2}
\end{subfigure}
\caption{\label{fig:LightCone}Real and imaginary parts of the eigenvalues of $S$ after contour rotations for fields and small time rotation. Hot colors correspond to larger eigenvalues.}
\end{figure}

\begin{figure}[ht]
\centering
\begin{subfigure}{.5\textwidth}
  \centering
  \includegraphics[width=.75\linewidth]{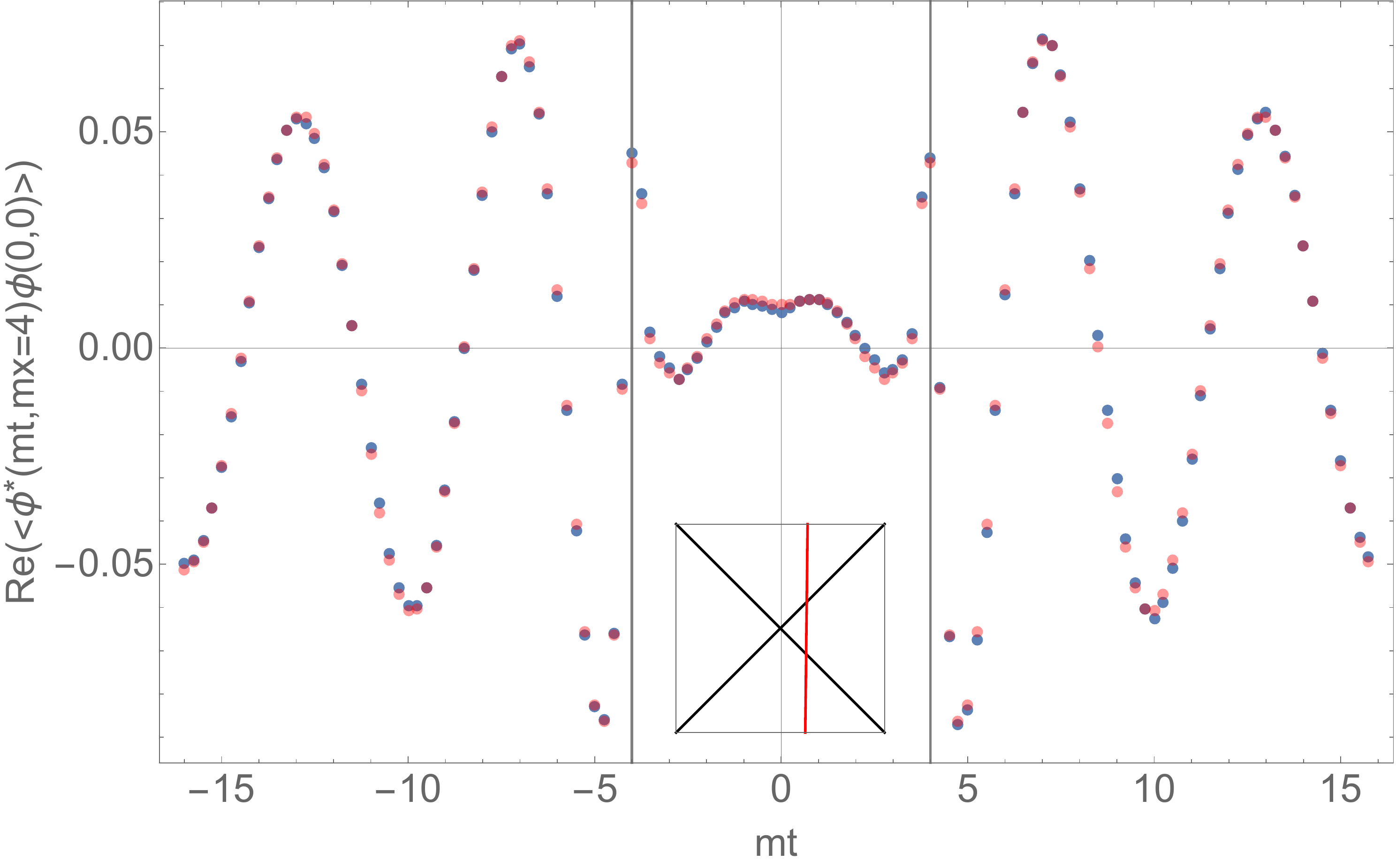}
  \label{fig:sub1}
\end{subfigure}%
\begin{subfigure}{.5\textwidth}
  \centering
  \includegraphics[width=.75\linewidth]{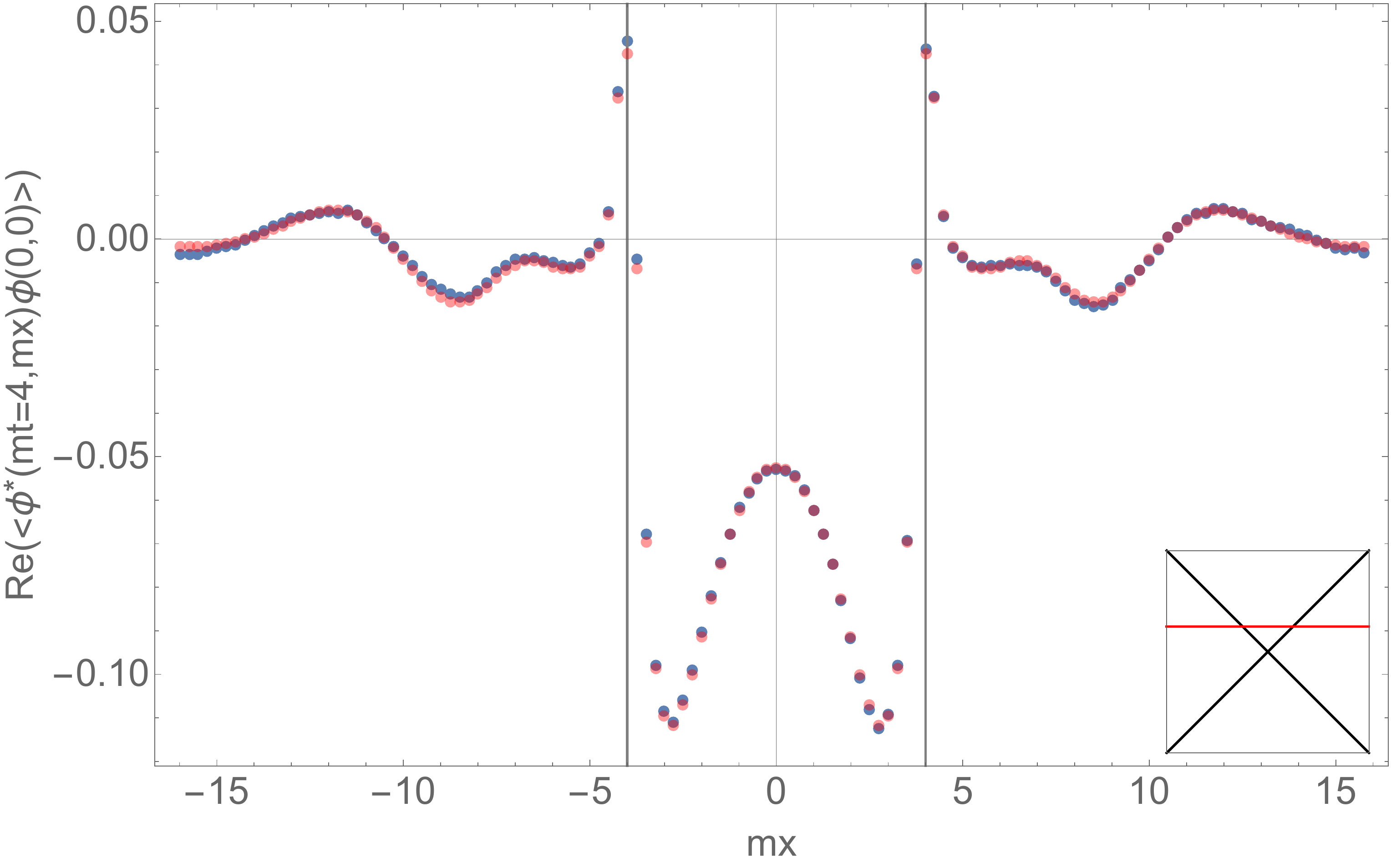}
  \label{fig:sub2}
\end{subfigure}
\caption{\label{fig:QFT_Propagators}Klein-Gordon propagator in $d=1+1$ evaluated by simulating the action (\ref{eq:QFT_rotated_action}).}
\end{figure}

The idea that we have presented in detail for the harmonic oscillator may be adapted to systems with infinitely many degrees of freedom without many additional difficulties. We start from the lattice action for a complex scalar in $d=1+1$:
\begin{equation}
S [ \phi ] = \frac{a^2}{2} \sum_{x, \mu} \left( \partial_{\mu} \overline \phi_x \partial^{\mu} \phi_x - m^2 \overline \phi_x \phi_x \right),
\end{equation}
where $\partial$ is the forward lattice derivative. Indices are raised and lowered with the Minkowski metric $\mathrm{diag}(1,-1)$. Fourier transformation followed by a rotation of inegration contours and inverse Fourier transformation leads to the action
\begin{equation}
i S[\psi] = - \frac{1}{2V} \sum_p \left| \hat p_{\mu} \hat p^{\mu} - m^2 \right| \left| \widetilde{\psi} (p) \right|^2,
\label{eq:QFT_rotated_action}
\end{equation}
where $\hat p^{\mu} = \frac{2}{a} \sin \left( \frac{a p^{\mu}}{2} \right)$. Fields are related by $\phi_x = \sum_y K_{xy} \psi_y$, $\overline \phi_x = \overline K_{xy} \psi_y^*$. We emphasize that after this transformation $\overline \phi$ is no longer equal to the complex conjugate of $\phi$. Similarly $\overline K_{xy} \neq K_{xy}^*$. Moreover the action is strongly nonlocal and UV singular. However we do have $e^{iS} >0$. There is still one problem to be fixed: modes corresponding to momentum close to the cutoff and near the light cone are not damped. Therefore any results obtained with this action are dominated by discretization errors. This is fixed by a small Wick rotation $x^0 \mapsto x^0 e^{- i \epsilon}$. The effect is that eigenvalues of $S$ corresponding to large momentum are damped. Below the cutoff this effect is negligible and the geometry of the light cone is accurately reproduced (cf. Fig. \ref{fig:LightCone}). Rotation of time introduces a mild sign problem which can be dealt with by reweighting. Using this technique we have evaluated the two point function. We present the results in the Figure \ref{fig:QFT_Propagators}. We observe oscillations inside and damping outside the light cone, which is a characteristic feature of Minkowski time propagators. We remark that we have seen finite volume effects much larger than in the Euclidean theory.

Transformation to (\ref{eq:QFT_rotated_action}) may be written down explicitly in the infinite volume continuum:
\begin{subequations}
\begin{gather}
\phi(x) = \int d^d y \left( e^{\frac{i \pi}{4}} \delta_+(x-y;m) + e^{- \frac{i \pi}{4}} \delta_-(x-y;m)  \right) \psi(y), \\
\delta_{\pm} (x ; m) = \int \frac{d^d p}{(2 \pi)^d} \theta ( \pm (p^2 - m^2)) e^{-ipx}.
\end{gather}
\end{subequations}
For $d=2$ this reduces to 
\begin{equation}
\left. \delta_+ (x;m) \right|_{d=2} = \frac{m}{2 \pi^2 \sqrt{- (x-i0)^2}}K_1 (m \sqrt{-(x-i0)^2}) + \mathrm{c.c.}.
\end{equation}
This distribution diverges quadratically as $x \to 0$. The action may also be written down as an integral over $x, y$ of a bilocal Lagrangian with a quadratic singularity for $x \to y$.

\section{Summary and outlook} 

We presented a treatment of the Beyond Complex Langevin approach clear of conceptual difficulties. In the process an interesting connection with thimbles has been observed. Moreover we have discovered that BCL actions considered so far have a hidden symmetry generated by differential operators $D, \overline D$. We hope that better understanding of this structure will allow to construct positive representations for more complicated complex weights in the future. As a side-effect of considering the negative eigenvalues of actions we developed techniques for simulating simple quantum systems in Minkowski time. Including interactions remains a challenge for the future.


\begin{thebibliography}{99}

\bibitem{Salcedo}
Salcedo, L. L.,
\emph{Representation of complex probabilities},
JMP 38 (97),
\emph{Existence of positive representations for complex weights},
JPA 40 (07).

\bibitem{Weingarten}
Weingarten, D.,
\emph{Complex probabiltiies on $\mathbb R^N$ as real probabilities on $\mathbb C^N$ and applications to path integrals},
PRL 89 (02).



\bibitem{RubaWyrzyk}
Ruba, B., Wyrzykowski, A.,
\emph{Explicit positive representation for complex weights on $\mathbb R^d$},
EPJ WC 175 (18).

\bibitem{Parisi}
Parisi, G.,
\emph{On complex probabilities},
PLB 131 (83).

\bibitem{Klauder}
Klauder, J.R.,
\emph{Coherent-state Langevin equations for canonical quantum systems with applications to the quantized Hall effect},
PRA 29 (84).

\bibitem{Pham}
Pham, F.,
\emph{Vanishing homologies and the $n$ variable saddlepoint method},
PSPM 40 (83).

\bibitem{Cristoforetti}
Cristoforetti, M. et al.,
\emph{New approach to the sign problem in quantum field theories: High density QCD on a Lefschetz thimble},
PRD 86 (12),
\emph{Monte Carlo simulations on the Lefschetz thimble: Taming the sign problem},
PRD 88 (13).

\bibitem{Wosiek_zzbar}
Wosiek, J.,
\emph{Beyond complex Langevin equations: from simple examples to positive representation of Feynman path integrals directly in the Minkowski time},
JHEP 04 (16); see also in these Proceedings.

\bibitem{AmbjornYang}
Ambjorn, J., Yang, S.-K.,
\emph{Numerical problems in applying the Langevin equation to complex effective actions},
PLB 165 (85).

\bibitem{Thesis}
Ruba, B.,
\emph{Badanie problemu znaku w mechanice kwantowej},
Master Thesis, UJ (18).

\bibitem{Duistermaat}
Duistermaat, J. J.,
\emph{On the Morse Index in Variational Calculus},
AM 21 (76).

\end{thebibliography}
\end{document}